\documentclass{article}
\usepackage{times}
\usepackage{fullpage}

\usepackage{amsmath,amsthm}
\usepackage{graphicx}
\usepackage{epsfig}
\usepackage{subfigure}

\begin{document}

\newtheorem{theorem}{Theorem}[section]
\newtheorem{lemma}[theorem]{Lemma}
\newtheorem{prop}[theorem]{Proposition}
\newtheorem{claim}[theorem]{Claim}
\newtheorem{corollary}[theorem]{Corollary}
\newtheorem{definition}[theorem]{Definition}
\newtheorem{assumption}[theorem]{Assumption}
\newtheorem{observation}[theorem]{Observation}
\newtheorem{example}[theorem]{Example}
\newtheorem{remark}[theorem]{Remark}
\newtheorem{algorithm}[theorem]{Algorithm}

\newenvironment{packed_itemize}{
\begin{itemize}
  \setlength{\itemsep}{1pt}
  \setlength{\parskip}{0pt}
  \setlength{\parsep}{0pt}
}{\end{itemize}}

\newcommand{\Xomit}[1]{}
\newcommand{\supproof}[1]{}

\renewcommand{\proof}[1]{
{\noindent {\it Proof.} {#1} \rule{2mm}{2mm} \vskip \belowdisplayskip}
}

\newcommand{\proofnoend}[1]{
{\noindent {\it Proof.} {#1} \vskip \belowdisplayskip}
}

\newcommand{\proofeq}[1]{
{\noindent {\it Proof.} {#1} }
}

\def\pfend{~~~~~~~~~~~\rule{2mm}{2mm}}

\newcommand{\rf}[1]{
{{\hspace*{-6pt} (\ref{#1}) \hspace*{-6pt}}}
}

\def\stm{}

\newcommand{\prevs}[2]{
{\vskip 0.1in \noindent {\em Proof of \rf{#1}.} {#2} \rule{2mm}{2mm}
\vskip \belowdisplayskip}
}

\newcommand{\prevproof}[2]{
{\vskip 0.1in \noindent {\bf Proof of {#1}.} {#2} \rule{2mm}{2mm}
\vskip \belowdisplayskip}
}

\def\rs{\vspace*{-0.1in}}

\newcommand{\xhdr}[1]{\paragraph*{{\bf #1}}}

\newlength{\saveparindent}
\setlength{\saveparindent}{\parindent}
\newlength{\saveparskip}
\setlength{\saveparskip}{\parskip}
\newenvironment{tiret}{%
\begin{list}{\hspace{2pt}\rule[0.5ex]{6pt}{1pt}\hfill}{\labelwidth=15pt%
\labelsep=5pt \leftmargin=20pt \topsep=3pt%
\setlength{\listparindent}{\saveparindent}%
\setlength{\parsep}{\saveparskip}%
\setlength{\itemsep}{2pt}}}{\end{list}}

\newenvironment{tiretx}{%
\begin{list}{\hspace{2pt}\rule[0.5ex]{6pt}{1pt}\hfill}{\labelwidth=15pt%
\labelsep=5pt \leftmargin=40pt \topsep=3pt%
\setlength{\listparindent}{\saveparindent}%
\setlength{\parsep}{\saveparskip}%
\setlength{\itemsep}{2pt}}}{\end{list}}

\def\eps{{\varepsilon}}
\def\Prf{{\rm Pr}}
\def\ev{{\cal E}}
\def\evf{{\cal F}}

\newcommand{\Prb}[1]{
\Prf\left[{#1}\right]
}

\newcommand{\Prg}[2]{
\Prf\left[{#1}~|~{#2}\right]
}

\newcommand{\Exp}[1]{
{\bf E}\left[{#1}\right]
}

\newcommand{\Expg}[2]{
E\left[{#1}~|~{#2}\right]
}

\newcommand{\pfrac}[2]{
\left(\frac{{#1}}{{#2}}\right)
}

\newcommand{\ofrac}[2]{
\frac{{#1}}{{#2}}
}

\def\subsetneq{\ \lower.5ex\hbox{$\stackrel{\subset}{\sneq}$}\ }

\newcommand{\set}[1]{\left\{#1\right\}}
\newcommand{\mnote}[1]{\marginpar{\footnotesize\raggedright#1}}
\newcommand{\brac}[1]{\left(#1\right)}
\newcommand{\bfrac}[2]{\brac{\frac{#1}{#2}}}

\def\agents{{\Omega}}
\def\O{{\cal O}}

\title{Information-Sharing and Privacy in Social Networks
}

\author{
Jon Kleinberg \hspace{.2in}
Katrina Ligett \\
Dept.\ of Computer Science\\
Cornell University, Ithaca NY\\
{\tt \{kleinber,katrina\}@cs.cornell.edu}
}

\date{}

\maketitle

\newcommand{\UnnumberedFootnote}[1]{{\def\thefootnote{}\footnote{#1}
\addtocounter{footnote}{-1}}}

\UnnumberedFootnote{
This work has been supported in part by
NSF grants CCF-0325453, BCS-0537606, IIS-0705774, 
IIS-0910664, 
CCF-0910940, 
a Google Research Grant,
a Yahoo!~Research Alliance Grant,
the John D. and Catherine T. MacArthur Foundation, and a NSF-funded Computing Research Association Computing Innovation Postdoctoral Fellowship.
}

\begin{abstract}
We present a new model for reasoning about the way information
is shared among friends in a social network, and the resulting
ways in which it spreads.
Our model formalizes the intuition that revealing personal 
information in social settings involves a trade-off between
the benefits of sharing information with friends, and the risks
that additional gossiping will propagate it to
people with whom one is not on friendly terms.
We study the behavior of rational agents in such a situation, 
and we characterize the existence and computability of stable
information-sharing networks, in which agents do not have
an incentive to change the partners with whom they share information.
We analyze the implications of these stable networks for
social welfare, and the resulting fragmentation of the social network.

\end{abstract}

\section{Introduction}
A growing line of work on privacy has investigated ways for 
people to engage in {\em transactions} --- purchases,
queries, participation in activities, and related types of
behavior --- while revealing very little or no private information
about themselves.
This research has implicitly construed the problem of privacy as
one of a trade-off between the concrete tasks that a person wants or needs
to accomplish, and the ``leakage'' of personal information that might
result from the interactions required to perform the task.
From such a framing of the problem, 
it follows that people should want to perform these tasks while
exposing as little information as possible.

If one takes this view of privacy, however, it becomes very hard to
reason about the kinds of simple, privacy-revealing activities that 
are ubiquitous in real social networks, both off-line and on-line.
As the most basic example, consider
two friends engaged in conversation, each sharing
personal --- though not necessarily particularly sensitive or important ---
information about themselves with the other: a child
is out sick from school; a scheduled trip was canceled;
some needed repairs on the house have just been finished.
Here, there is no transaction taking place other than the
sharing of the information itself, and it is easy to create
scenarios in which any of these seemingly mundane pieces of information
could ultimately be used to the detriment of the person revealing it.
Yet in everyday life people clearly feel a fundamental incentive to engage in 
this kind of information-sharing;
if we are to understand the full scope of privacy as an issue,
we need to be able to model and reason about this kind of activity
with the same level of concreteness that we use for 
on-line purchases, database and search-engine queries,
and the other more formal, structured types of transactions that
have been the traditional focus of privacy research.

\xhdr{Information-Sharing in Social Networks}
As the first step toward developing a model for this kind of
activity, it is useful to try articulating some aspects of 
the unstated social conventions that govern the informal sharing of
personal information between two friends.
This question touches on complex issues from several research literatures,
including sociology, psychology, and legal philosophy,
and the work on this topic has elucidated both positive and
negative aspects of information-sharing practices
\cite{adams-friendship,aries-friendship,bazerman-negotiation,
reiman-friendship}.
Given this complexity, we will try to abstract some of
the most basic aspects of information-sharing in social networks
into a mathematically tractable model.
In particular, if we want to explore the potential for rational agents
to engage in information-sharing with friends, 
we need to formalize sources of positive utility that derive
from this activity, to trade off against the 
the sources of negative utility that have been the 
dominant focus in the computer science literature on privacy.

With these issues in mind, and drawing on the literature above,
we argue that personal conversations 
between you and a friend are governed by social conventions
that, at a general level, contain the following general ingredients.
\begin{itemize}
\item[(i)] You derive benefit from learning information about your friend, 
in part because such exchanges serve to strengthen the social
tie between the two of you.  Moreover, there is a corresponding benefit
in having your friend learn information about you; this too
strengthens the social tie.
\item[(ii)] It is unrealistic to have all such conversations governed
by strict promises of secrecy; all parties involved can expect that
some information will spread through the social network
to a limited extent via gossip.
\item[(iii)] The fact that information spreads through the social
network contains sources of both positive and negative utility for you.
You may receive positive utility from learning information about friends and
having friends learn information about you, even by this form of indirect
transmission through gossip.
However, there are other people in the network whom you do not want
your personal information to reach; you receive negative utility
when personal information about you indirectly reaches them via gossip.
\item[(iv)] In evaluating whether to share personal information 
with a friend, you therefore take into account who else you believe this
friend engages in information-sharing with --- and more generally, what
you believe the information-sharing pathways in the network look like.
You will avoid sharing information with a friend if you believe that
their indirect transmission of your information will yield a net negative
utility.  Correspondingly, you may avoid sharing information with 
a friend if by cutting this link, it will encourage others to feel
safe in sharing their information with you --- provided that this trade-off
yields a net utility benefit for you.
\end{itemize}

These general considerations form the basis for the model we develop next.
There are many further, and important, 
issues that could be incorporated into a model:
for example, information comes in many categories, and
you may well be happy if person $X$ learns about your personal information
related to topic $Y$ but not to topic $Z$;
similar contrasts may exist when you consider your personal information
classified not by topic but by its level of sensitivity.
However, we will see that building a model even from the most
basic considerations above already leads to complex questions,
with results that provide insight --- and appear to accord with
natural intuitions --- about some of the ways in which personal information
 moves through social networks.

\xhdr{Formalizing a Model of Information-Sharing}
We now describe a model that takes into account issues (i)-(iv)
from the preceding discussion.
We begin by describing a model without any strategic component
on the part of the people involved, and then we add a strategic
aspect to it.

We have a set $V$ of $n$ people; some pairs of these people share
personal information with each other (including any indirect information
that they've learned about others), and some pairs of these people 
do not share personal information.
Sharing of information is symmetric, and so if we let $E$ denote
the set of pairs who share information, then we obtain an
{\em information-sharing network} $G = (V,E)$.
If $i, j \in V$ are in the same connected component of $G$, then each will
learn personal information about the other, either by direct
communication (if there is an $i$-$j$ edge) or indirectly via
gossip (if there is only an $i$-$j$ path of length two or more).

Now, for for any two people $i, j \in V$,
person $i$ receives a utility $u_{ij}$ from being in the same
component as $j$, and person $j$ receives a utility $u_{ji}$
from being in the same component as $i$.
These utilities can be either positive or negative, corresponding 
to the dichotomy in point (iii) above between the benefits of
indirectly learning about and being known to your friends, and the harms 
from having personal information reach people you are not friendly with.
If $C_G(i)$ denotes the component of $G$ containing $i$, then
the total utility of $i$ is equal to 
$\sum_{j \in C_G(i)} u_{ij}$.

\xhdr{Strategic Behavior and Information-Sharing}
In our model, two people must mutually agree to share information,
and we presume that they will do so strategically, to maximize
their utilities, based on their expectations about what others will do.
This is the crux of point (iv) in the preceding discussion.
(We think of the set $V$ as being a relatively small community,
such that everyone has beliefs about who is talking to whom.)

We are thus faced with a kind of network formation game, in which 
each player must decide which links to maintain so as to maximize
her utility, given the links everyone else has formed.
We seek information-sharing networks that satisfy a type of stability;
from a stable network $G$ there will be 
no incentive for parties to add or drop links.
In other words, people will be sharing information with the
optimal set of contacts, given the pathways for gossip formed
by the behavior of everyone else, and they can trust that
the information-sharing structure that has developed is thus
in a sense ``self-enforcing.''
Note also that in contrast to many standard network formation games,
there is no explicit ``cost'' to maintain a link; the costs
are implicit, based on the fact that a link exposes you to the risk
that your information will reach other nodes with whom you are not friendly.

Our stability notion is a strengthening of {\em pairwise Nash stability}
\cite{jackson-networks-book}
(see the next section for a review of alternate stability notions).
Specifically, we define a {\em defection} from the current network $G$
to consist either of (a) a single node $i$ deleting a subset of its incident
edges, or (b) a pair of nodes 
$i, j$ agreeing to form the edge $(i,j)$ and simultaneously to each
delete subsets of their (other) incident edges.
We then say that a graph $G$ is {\em stable} if there are no defections 
from $G$ in which all the participating nodes ($i$ in case (a), and both $i$
and $j$ in case (b)), strictly improve their utilities.
From the results that follow, it will become clear that defining
a network to be ``stable'' without allowing two-node coordination
of the type in (b) provides a stability concept that is too weak
to reflect the strategic information-sharing behavior 
we are trying to capture.\footnote{Note that for defections of type (b),
we require both $i$ and $j$ to strictly improve their utilities.
This is in keeping with an assumption that utilities are not transferable
(so that e.g. $i$ cannot pay $j$ to join her in a defection),
and we will see that it creates a theoretical framework that
more naturally connects to related lines of work in 
strategic network formation.}

The ability of two people to coordinate is natural in our model,
since pairwise interaction is the fundamental level at which
information-sharing is taking place.
But it is also interesting to consider the possibility of defections
in which larger subsets of people coordinate their actions.
Thus, we define a {\em $k$-defection} to consist of a set $S$
of up to $k$ nodes agreeing to form all pairwise edges within $S$,
and simultaneously to each delete subsets of their (other) incident edges.
We say that $G$ is {\em $k$-stable} if no $k$-defections are
possible from $G$.\footnote{One could also consider a notion of
$k$-defection in which the $k$ nodes in $S$ only form a subset of
the edges within $S$.  Since we want to capture the idea that the
whole set is mutually coordinating, rather than consisting of two disjoint
sets that act simultaneously, we adopt the definition in which 
all edges are formed.  We also note that a variant of the definition
in which a connected (but not necessarily complete) subgraph on $S$ is formed 
yields very similar lines of analysis, since nodes' utilities are
derived from the components they belong to, rather than just who
they are directly connected to.}
In view of this general definition, we will sometimes refer to
the defections and stability notion in the previous paragraph
as {\em $2$-defections} and {\em $2$-stability}.

As noted above, there are many possible generalizations of this model.
For example, there could be different categories and 
different sensitivities of information;
information could ``attenuate'' as it travels over multi-step paths,
perhaps being forgotten with some probability at each step;
and our model does not include the notion of globally ``publishing''
personal information through a mechanism like a personal Facebook page,
but instead focuses on person-to-person communication.
Thus, we can think of the model as capturing the information-sharing
relationships for a single kind of information, by direct interaction,
and in a coherent enough community that people have expectations
about the behavior of others.
Extending these assumptions in any of the above directions would
be an interesting focus for future work.

\xhdr{Our Results: Existence and Social Welfare}
Our central goal is to study the most basic version of the model 
that is still rich enough to yield non-trivial and meaningful outcomes.
Thus, for much of the first part of the paper, we focus on
the case in which utilities are symmetric ($u_{ij} = u_{ji}$)
and take values from the set $\{-\infty,1\}$.
This corresponds to a natural version of the problem in which
all pairs of people are either {\em friends} or {\em enemies};
there is a positive utility in sharing information with friends,
but a much stronger negative utility in having enemies find
out information about you.

Our first main result is that for any set of symmetric utilities from
$\{-\infty,1\}$, and every $k \geq 2$, a $k$-stable network always exists.
For $k = 2$ --- the basic definition of stability --- we can find
a stable network in polynomial time.
For general $k$, it is NP-hard to construct a $k$-stable 
network.\footnote{In other words, although the decision problem, ``Does there
exist a $k$-stable network?'' has the trivial answer ``yes,''
an algorithm that produces a witness could be used to solve NP-complete
problems.}
The intermediate case of fixed, constant $k > 2$ is interesting;
we show how to construct $k$-stable networks in polynomial time 
for $k = 3$ and $k = 4$, with larger constants $k$ left as open questions.

There are also natural questions related to 
the notions of {\em social welfare,} defined
as the sum of utilities of all nodes, and 
{\em socially optimal networks,} defined as those that maximize
social welfare.
Since a socially optimal network may not be stable, 
we can ask about the {\em price of stability} --- the maximum welfare of 
any stable network relative to the optimum.
We find that the price of stability is equal to $1$ for
2-stable and 3-stable networks --- in other words, there always
exist such networks achieving the social optimum --- 
but it exceeds $1$ for $k > 3$.
It is an open question to find a tight bound on the price
of stability for $k > 3$.

\xhdr{Our Results: Connections to Graph Coloring}
There is a natural connection between the case of 
symmetric utilities from $\{-\infty,1\}$ and the problem
of {\em graph coloring}. 
Indeed, if we let $F$ denote the pairs
of nodes $(i,j)$ with utility $u_{ij} = -\infty$, and 
define the {\em conflict graph} for the instance of
the problem to be $H = (V,F)$, then
the components in any stable network $G$ will have to be
independent sets of $H$, and hence correspond to a coloring of $H$.
The requirements of stability, of course, demand more,
and so we in fact get an interesting and novel variant of
the graph coloring problem in which we must find a coloring 
in which nodes in different color classes are all
``blocked,'' in a certain sense,
from wanting to form direct connections with each other.

Using the connection to graph coloring, we can consider the following
alternate definition of welfare for an information-sharing network $G$:
the number of components it has.
This essentially captures the extent to which nodes' collective avoidance of
information leakage has caused the group to ``fragment'' into
non-interacting components.
It is natural to want this number of components to be as small as possible,
relative to the minimum achievable if we did not require stability;
this minimum is $\chi(H)$, the chromatic number of the conflict graph $H$.
We show that there is always a 2-stable network with a number
of components equal to $\chi(H)$, and hence
the analogue of the price of stability is equal to $1$ when
the number of components is used to measure welfare.
On the other hand, when we consider $n$-stable networks --- the extreme
case in which we allow defections of arbitrary size --- it can
be the case that the only $n$-stable networks have a number of
components equal to $\Omega(\log n) \cdot \chi(H)$; and we show
that this bound is tight, by proving that there is always 
an $n$-stable network with at most $O(\log n) \cdot \chi(H)$ components.

\xhdr{Our Results: General Forms of the Model}
Let's now return to the general formulation of the problem, in which 
for each pair of nodes $i$ and $j$, node 
$i$ receives a utility $u_{ij}$ from being in the same
component as $j$, and we may have $u_{ij} \neq u_{ji}$.

It turns out that many problems involving notions of stability
or self-enforcing relations are contained in this general version.
For example, the Gale-Shapley {\em Stable Marriage Problem} 
with $n$ men and $n$ women \cite{gale-shapley-stable-marriage}
arises as a simple special case of the model,
by defining $u_{ij} = -\infty$ for each pair of men and
each pair of women, and when a person $i$ has a person $j$ of
the opposite gender in position $p$ on his or her preference list, 
defining $u_{ij} = 1 + n - p$.
Related problems such as Becker's {\em Marriage Game} 
\cite{Becker73,Becker74} can be
similarly reduced to simple forms of the present model.

One downside of this generality is that once we move even
a little beyond the case of symmetric utilities in $\{-\infty,1\}$,
the problem quickly becomes intractable.
In particular, consider a case that is just slightly more general:
symmetric utilities from $\{-\infty,1,n\}$.
In other words, the friendly relations now
consist of ``weak ties'' of weight $1$ and 
``strong ties'' of weight $n$ \cite{granovetter-weak-ties},
with the relative values chosen
so that the benefit of a single strong tie outweighs the
total benefit of any number of weak ties incident to a single node.
We show by a simple example that stable networks need not always
exist with these kinds of weights.\footnote{The simplest 
such example has four people:
Anna has a strong tie to Bob; Claire has a strong tie to Daniel;
Bob and Daniel are enemies; and all other relations are weak ties.
In any stable network, Anna and Bob would need to belong to the same
component; Claire and Daniel would need to belong together in a 
different component.  But then Anna and Claire would have an incentive
to form an edge, violating stability.}
More strongly, we show in fact that for any $k$, 
deciding whether a given instance contains a $k$-stable network 
is NP-complete; the proof of this is based on developing the
connection with graph coloring more extensively.

Despite this hardness result, the presence of elegant special cases
like the Stable Marriage Problem suggests that there is considerable
promise in developing a deeper understanding of the structural conditions
on utilities that lead to settings in which stable networks always
exist, and in which they can be efficiently identified.

\xhdr{Organization of the Paper}
The remainder of the paper is organized as follows.
In Section~\ref{sec:related}, we discuss the connections between
our model and related work in economic theory and computer science.
In Sections~\ref{sec:existence}, \ref{sec:utility}, 
and \ref{sec:components}, we discuss
our results on the existence, efficient construction, and
welfare properties of stable networks for symmetric utilities 
in $\{-\infty,1\}$.
Finally, in Section~\ref{sec:general}, we discuss our results on
more general forms of the model.

\section{Related work}
\label{sec:related}

Closely related to this work is the substantial literature in economics on coalition formation.  Coalitional games model the partitioning of a society into collaborative groups that jointly create worth; the value of each group is then shared among its participants.  We restrict our discussion here to models involving non-transferable utility, which excludes, for example, the work of Deng and Papadimitriou 
\cite{DP94}.  It also distinguishes our work from that of Muto \cite{Muto90} and Nakayama and Quintas \cite{NQ91}, which further differs ours in that their model does not incorporate network structure and uses a different definition of stability.  Existing work on coalition formation differs from our work in two substantial ways:  First, the solution concepts and defection models used in the coalition formation literature are fundamentally not suitable for modeling gossip and information leakage in a social network.  Second, much of the foundational work in coalition formation focuses on conditions for the \emph{existence} of these orthogonal solution concepts, whereas we study not only existence but \emph{computational issues} and consequences for \emph{social utility}.

\xhdr{Solution Concepts Requiring Large-Scale Consensus}
Almost all solution concepts for coalitional stability require that
outcomes be \emph{individually rational}, meaning no player would get
higher utility by being alone in a singleton coalition.  Beyond this,
though, much of the work on solution concepts for coalitional
stability pertains to definitions of deviations that require the
consensus of a large number of players.  We contend that
information spreads in a social network not only by centralized
dissemination strictly within globally negotiated coalitions, but
through relationships negotiated at a local scale by a small number of
individuals without the permission of the group as a whole.  

The best
established solution concept in the coalitional literature is the
\emph{core}, which consists of player partitions and payoff
distributions so that there is no subset of the players that all are
willing to simultaneously abandon their current coalitions and form a
new one (where ``willing'' means at least one of the deviating players
must strictly prefer the deviation).
There is a substantial literature on necessary and sufficient
conditions for the non-emptiness of the core, including work by
Bannerjee et al.~\cite{BKS01} and Bogomolnaia and Jackson \cite{BJ02}.

Bogomolnaia and Jackson \cite{BJ02} also study conditions on player preferences
that imply existence of \emph{individually stable coalition
  partitions}, which model individual player defections by requiring
that every player in the coalition a defecting player wishes to join must
agree to the defection.  Again, this type of global coordination is
not a good model for gossip: inherent in the concept of gossip is that
it spreads without the permission or knowledge of the individuals to
whom it pertains.

Like us, Dimitrov et al.~\cite{DBHS06}
consider games where players have friend or enemy
relationships; they characterize the \emph{internally stable coalitions}, where no subgroup of any coalition wishes to break off and form a new coalition.  While internal stability under small group defections might be a reasonable criterion for stability of a gossip network, this solution concept doesn't allow for the possibility that two players from different coalitions might benefit from pooling their information.  Dimitrov et al.~\cite{DBHS06} and Elkind and Wooldridge\cite{EW09}  both also study the computational tractability of computing the core.

Barbera and Gerber \cite{BG07}
observe that
no solution concept can simultaneously
provide a number of desirable properties.
Among other things this argument ignores the difficulty of
coordinating a defection by a large number of players.

\xhdr{Nash Stability}
The concept of \emph{Nash stability}
comes closest to our defection model; it
describes situations where no player
wishes to unilaterally defect to join a different coalition
(regardless of whether they would agree to receive her).  For our
purposes, however, this is too individualistic: the spread
of information should require the participation of at least two players.
Milchtaich and Winter \cite{MW02} use the Nash stability concept, and
study a model where players prefer to associate with other players who
are similar to them, but there is some upper bound on the total number
of groups allowed.  Here, Nash-stable partitions might not exist.
In addition to studying existence,
they also are interested in distributed equilibrium computation: they
show that asynchronous myopic randomized better response converges
almost surely to a stable partition (under a somewhat limited
definition of better response, where defecting players do not account
for the impact they would have on the coalition they are joining).

\xhdr{Social Welfare}
Branzei and Larson
\cite{BL09} consider a model that is similar to ours, where each agent has a value for being in the same coalition as each other agent and utility is non-transferable.  They also consider issues of social welfare, but for stability concepts (the core, internal stability) unsuitable for the study of information-sharing.

\section{Existence and computation of stable outcomes}
\label{sec:existence}

\subsection{$2$-stability}

We begin with the most basic model described in the introduction;
we consider 2-stable networks for the case of symmetric utilities
from $\{-\infty,1\}$.

We first show that the following (inefficient) algorithm always produces a
2-stable network $G = (V,E)$.  
Recall that the {\em conflict graph} $H = (V,F)$ is a graph on the
same node set as $G$, 
defined by setting $F = \{(i,j) : u_{ij} = -\infty\}$.
\begin{algorithm}\label{alg:undir-def3}
\mbox{~}\\
\begin{packed_itemize}
\item Find a maximum-size independent set $S$ in $H$.
\item Add all pairwise edges on $S$ to the graph $G$.
This clique on $S$ will be one of the components of $G$.
\item Iterate on $V - S$.
\end{packed_itemize}
\end{algorithm}

\begin{theorem}
Algorithm \ref{alg:undir-def3} produces a 2-stable network.
\label{thm:2-stab-exist}
\end{theorem}

\proof{First, because all components of $G$ are built from independent
sets of $H$, all nodes have non-negative utility in $G$.
Thus, no node wants to defect by unilaterally deleting incident edges.

Now suppose that there were a defection in which two nodes $i$ and $j$
wanted to form the edge $(i,j)$, potentially deleting some of their
incident edges.
Let $I$ denote the component of $G$ containing $i$,
and let $J$ denote the component of $G$ containing $j$.
Suppose (by symmetry) that $I$ was formed by the algorithm 
before $J$.
Then there is some $i' \in I$ for which $(i',j) \in F$, since
otherwise $j$ could have been included in $I$ when $I$ was formed.

Since $i$ has edges to all nodes in $I$, a defection by $i$ and $j$
will only be utility-increasing for $j$ if $i$ deletes all its
incident edges.
Given this, a defection by $i$ can only be utility-increasing
if (a) $j$ retains edges to nodes in $J$, (b) $i$ has no $-\infty$-edge to any
$j' \in J$, and (c) $|J| + 1 > |I|$.
But in this case, $J \cup \{i\}$ is an independent set in $F$
of cardinality strictly greater than $I$, which means that
in the iteration when the algorithm constructed $I$,
it should have constructed $J \cup \{i\}$ instead.
}

We note that the use of maximum-cardinality independent sets
is crucial for this algorithm; the variant that
repeatedly identifies and deletes inclusionwise maximal independent
sets in $H$ need not create a 2-stable network.

Given that this algorithm contains the NP-hard maximum independent set
problem as a subroutine, we next consider the question of finding
a 2-stable network efficiently for any set of 
symmetric utilities in $\{-\infty,1\}$.
One approach is to consider iterating the analogue of best-response
dynamics for 2-defection: we repeatedly search for a 2-defection
from the current graph, and if we find one we have the node or nodes
perform the defection that maximizes their improvement in total utility.

Unfortunately, best-response dynamics can cycle
indefinitely, as we now show.

\begin{figure}
\begin{center}
\includegraphics[width=5cm]{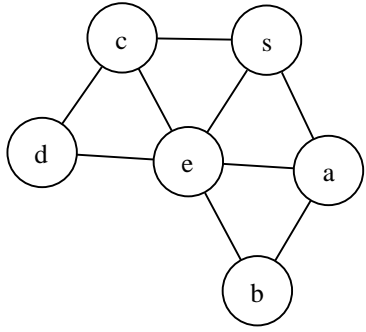}
\end{center}
\caption{
Depicted edges $(i,j)$ represent $u_{ij} = 1$; absent edges have $u_{ij} = -\infty$.}
\label{fig:best-response-cycles}
\end{figure}

\begin{theorem}
Best-response dynamics can cycle, with symmetric utilities 
in $\{-\infty, 1\}$.
\end{theorem}

\proof{
As a starting graph $G$, we take a large clique $s$ 
with identical utilities, plus five additional nodes 
$a, b, c, d, e$, depicted in Figure \ref{fig:best-response-cycles}.

Throughout the following best-response trajectory, 
the nodes of $s \cup \{e\}$ will remain in a clique.
The starting network $G$ will also contain the two additional edges
$(a, e), (c, d)$, and we make best-response moves as follows:

\begin{enumerate}
\item Change to $(c, e), (c, d)$ because $e$ and $c$ make a move: they form an edge and $e$ drops $a$.
This is a best response for $e$, who could have connected to $b$ (equally
good) or $d$ while dropping $a$ (also equally good).  This is a best
response for $c$, who had no other options.
\item Change to $(c, e)$ because $d$ drops its connection to $c$.  This is a best response for $d$.
\item Change to $(a, b), (c, e)$ because $a$ and $b$ form a link.  This is a best response for
each of them.
\item Change to $(a, e), (a, b)$ because $a$ and $e$ make a move: they form an edge and $e$ drops $c$.
The argument parallels that in step 1.
\item Change to  $(a, e)$ because $b$ drops its connection to $a$.  The argument parallels that in step $2$.
\item Change to $(a, e), (c, d)$ because $c$ and $d$ form a link. The argument parallels that in step 3.
\end{enumerate}

We've now returned to the initial network $G$, completing the proof.
}

Despite this cycling behavior, we now show how to perform 
a natural alternate dynamic process that reaches a stable network
in polynomial time.

\begin{theorem}
We can find a $2$-stable network in polynomial time.
\label{thm:2-stab-efficient}
\end{theorem}

\proof{
We build up a polynomial-length
sequence of networks iteratively, ending at a 
2-stable network.
We start from the network in which each node forms its own component,
and we inductively maintain the property that all 
intermediate networks in the sequence 
will have connected components consisting of cliques.

At each intermediate state, we look for a node $j$ in a clique $J$, such
that there is some clique with $|I| \geq |J|$, and no edge
$(i,j) \in F$ for any $i \in I$.
If we cannot find such a node, then the network is 2-stable, by
an analogue of the argument in the proof of Theorem~\ref{thm:2-stab-exist}.
Otherwise, we delete all of $j$'s edges to $J$, and create edges from
$j$ to all nodes in $I$.
(Note that this is not a 2-defection, but we are not producing
a run of best-response dynamics, simply a sequence of networks.)

Note that if there is an improving defection, there must be some node $j$ in a clique $J$, such
that there is some clique with $|I| \geq |J|$, and no edge
$(i,j) \in F$ for any $i \in I$.  By construction, no node wishes to unilaterally drop all her edges, and no two non-singleton nodes improve their utility by forming an edge between them while both dropping all of their edges.  If two nodes wish to form an edge while one of them drops all of her edges, this node is such a $j$.  If two nodes wish to form an edge while neither drops all her edges, the node from the smaller (or either, if the cliques are of equal size) clique provides such a $j$.

Thus, our sequence of networks proceeds by repeatedly moving a node $j$
from one clique $J$ into another $I$, such that $|I \cup \{j\}| > |J|$.
We now show that this process must terminate after passing through
at most a polynomial number of networks.
For this, we let $x_0, x_1, x_2, \ldots$ denote the sizes of 
the cliques in our current graph, and we consider the potential function
$\sum_i x_i^2$.
If a player moves from a clique of size $b$ to a clique of size $a \geq b$,
then in the potential function we replace the terms $a^2 + b^2$ by
$(a+1)^2 + (b-1)^2 = a^2 + b^2 + 2(a-b) + 2 \geq a^2 + b^2$.
Thus, the potential function increases by at least $2$ with each move,
and since it can't grow larger than $n^2$, this proves that
the construction terminates after passing through at most $O(n^2)$ graphs.

The running time of the full algorithm is also polynomial, since we can
easily check for the existence of the required node $j$ in each 
iteration in polynomial time.
}

\subsection{$k$-stability}

We now consider the generalization to $k$-defections and 
the corresponding notion of $k$-stability.
We begin by showing that $k$-stable networks exist, for all $k$.

\begin{theorem}
For every $k \geq 2$,
every instance admits a $k$-stable network.
\end{theorem}

\proof{
In fact, we show that 
Algorithm~\ref{alg:undir-def3} finds a network that
is $k$-stable for all $k$.

Suppose by way of contradiction that in the network $G$
produced by this algorithm (consisting of disjoint cliques),
there were a set $S$ of nodes that wanted to defect.
Consider the first clique $I$ in order of formation that contains
a node $i \in S$.
All nodes in $S - I$ must have $-\infty$-edges to nodes in $I$,
so in any defection involving $S$, the node $i$ must drop all its
edges into $I$.

Now, let $I'$ be the component that $i$ belongs to after the defection.
In order for this to be a defection in which $i$ participates, it
must be that $|I'| > |I|$;
but then in the iteration when $I$ was produced, the algorithm
should have produced $I'$ instead, a contradiction.
}

However, although $k$-stable networks must exist,
actually constructing one is NP-hard.
\begin{theorem} Constructing a $k$-stable network 
is NP-hard when $k$ is part of the input.
\end{theorem}

\proof{
If $k$ is at least the size of the maximum independent set in the
graph $H$, any $k$-stable network contains a
maximum independent set of $H$ as one of its connected components.
}

In fact, 
even deciding whether a given network is $k$-stable 
is computationally intractable.
\begin{theorem} 
Testing stability under $k$-defections is NP-hard.
\end{theorem}

\proof{
The proof is by reduction from finding a $k$-node independent set.
Given an $n$-node graph $L$ that is an instance of independent set,
assume that each edge in the graph represents a $-\infty$ relationship
and that all absent edges are $+1$ relationships.  We will add $k-2$
additional nodes for each node $x_i$.  The $k-2$ nodes for $x_i$
all have $+1$ relationships with each other and with $x_i$ and have
$-\infty$ relationships with all other nodes in the graph.  The
arrangement whose stability we will test consists of $n$ many
$(k-1)$-node cliques,
each consisting of a node in the original graph and its $k-2$
additional nodes.  There is a group of $\leq k$ players who wish to
defect from this arrangement if and only if there was an independent
set of size $k$ in $G$.
}

Now, a natural question is whether it is computationally feasible
to construct $k$-stable networks for constant $k$.
One approach to this is to follow the style of analysis in 
the proof of Theorem~\ref{thm:2-stab-efficient},
and to use a potential function on the vector of component sizes
that always increases, and is bounded by a function of
the form $n^{f(k)}$.
Here something interesting happens: this approach provides a
polynomial bound when $k \in \{3, 4\}$, but we show that
such a cardinality-based potential function 
provably cannot provide a polynomial bound when $k \geq 5$.

To give some first intuition 
for what goes wrong, suppose we were to try using 
the function $\sum_i x_i^k$, where the $x_i$ are the component sizes.
Now, suppose $k = 6$; we 
consider 5 groups of 5 nodes,
and one group with 1 node; and we allow six nodes to defect.
Suppose further that we have have
one player from each large group all join the group of 1.  The initial
potential was $5^7 + 1 = 78126$ and the new potential is $5 \cdot 4^6
+ 6^6 = 67136$.  

We now provide proofs for the cases of $k \in \{3, 4\}$ and $k \geq 5$.

\begin{theorem}
We can find a $3$-stable network in polynomial time, 
using the potential function $\sum_i (x_i + 4) (x_i-1)/2$.
\end{theorem}

\proof{
We define a recurrence relation for a potential function for $k =3$:
\begin{align*}
F_3(1) &= 1\\
F_3(2) &= 3\\
F_3(3) &= 7 \\
F_3(i) &= 2 F_3(i -1) - F_3( i - 2) + 1
\end{align*}
gives $F_3(n) = (n + 4) (n-1)/2$.

In general, we require that $F_3(i) < F_3(i + 1)$.

The recurrence for $2$ nodes each leaving groups to form a new group
requires $2F_3(1) < F_3(2)$.  The recurrence for $1$ node leaving
its group to join another is 
\[F_3(i) = 2 F_3(i-1) - F_3(i -2),\]
which is strictly less than that given above.

This recurrence covers the worst case for $2$ nodes each leaving a separate
group and joining a third node.
The recurrence for $2$ nodes both leaving the same group and joining
a third node is
\[F_3(i) = F_3(i -1) + F_3(i - 2) - F_3(i - 3) - 1,\]
which is strictly less.
The recurrence for $3$ nodes each leaving groups to form a new group
requires that $3F_3(2) < 3 F_3(1) + F_3(3)$.
}

\begin{theorem}
We can compute a $4$-stable network in polynomial time, using a potential function that is $O(n^3)$.
\end{theorem}

\proof{
We will solve a recurrence relation to derive a  potential function for $k =4$:
\begin{align*}
F_4(1) &= 1\\
F_4(2) &= 3\\
F_4(3) &= 7\\
F_4(4) &= 17\\
F_4(i) &= 3F_4(i-1) - 3 F_4(i - 2) + F_4(i-3) + 1
\end{align*}
which solves to $F_4(n) = 17 (n-3) + (n-5)(n-4)(n-3)/6 + 7(n-5)(n-4)/2$ for $n \geq 6$.

In general, we require that $F_4(i) < F_4(i + 1)$.

Note that $F_4(i) \geq F_3(i), \forall i$, and thus 
we need only address defections by $4$ nodes.
The given recurrence covers the worst case 
for $3$ nodes each leaving groups and
joining a fourth node: If $3$ nodes left the same group to join a fourth, the recurrence is $F_4(i) > F_4(i -1) + F_4(i -2) - F_4(i - 3)$.  If $2$ nodes leave one group and one node leaves another, the recurrence is $F_4(i) > 2 F_4(i -1) - F_4(i-4)$.
The recurrence for $4$ nodes each leaving groups to form a new group
requires that $3 F_4(3) < 3 F_4(2) + F_4(4)$.
}

Starting with $k=5$, however, polynomially bounded additive potential functions no longer exist.
\begin{theorem}
Any additive potential function for $k$-defections on $n$ nodes for $k \geq 5$ is in $\Omega(2^n)$.
\end{theorem}

\proof{
We will lower bound the value of any potential function $F$ for $5$-defections.  First, let $F(1) = 1$.  Note that $2 F(1) < F(2)$ in order to increase the potential when $2$ singleton nodes defect to form a group.  In order to increase the potential when $3$ nodes defect from groups of size $2$ to form a new group, $3 F(2) < 3 F(1) + F(3)$.  Similarly, we get $4F(3) < 4F(2) + F(4)$ and $5F(4) < 5 F(3) + F(5)$.  Solving, this gives $F(2) \geq 3, F(3) \geq 7, F(4) \geq 17, F(5) \geq 51$, and thus $F(i) \geq 2^{i-1}$ for $i \leq 5$.

We now consider defections where $k-1 = 4$ nodes each defect from groups of size $i -1$ to join a group of size $i - k + 1 = i -4$, resulting in a group of size $i$ and $4$ groups of size $i - 2$.  Thus, $F(i) > 4 (F(i-1) - F(i-2)) + F(i - 4)$ for $i > 5$.  So certainly $F(i) \geq 4(F(i-1) - F(i-2))$ for $i > 5$, which solves to $F(i) \geq 2^{i -1}$.
}

\section{Social Welfare: Total Utility}
\label{sec:utility}

As noted in the introduction, there are two natural measures of
welfare for an information-sharing network $G$: the sum of node utilities,
and the number of components of $G$.
We first observe that optimizing each of these (over all networks,
not just stable ones) is NP-hard.

\begin{observation}
Maximizing the total utility on a $3$-partite graph is equivalent to partitioning the graph into induced triangles, which is NP-hard.
\end{observation}

\begin{observation}
Minimizing the number of groups in a partition is equivalent to determining the chromatic number of the graph, and thus cannot be approximated to within $n^{1 - \epsilon}$ for any $\epsilon > 0$.
\end{observation}

\begin{figure}
\begin{center}
\includegraphics[width=5cm]{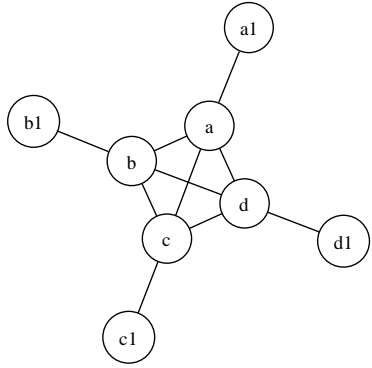}
\end{center}
\caption{
Depicted edges $(i,j)$ represent $u_{ij} = 1$; absent edges have $u_{ij} = -\infty$.}
\label{fig:chromatic-number-vs-total-util}
\end{figure}

The two notions of welfare are also quite distinct: 
networks that are ideal for one may not be optimal for the other.
\begin{theorem} There exist instances where no network
minimizing the number of connected components also maximizes the total utility.
\end{theorem}
\proof{ Figure \ref{fig:chromatic-number-vs-total-util} has only one
network that minimizes the number of connected components while not placing
any $-\infty$-edges within a connected component: the four pairs $\{a,
a1\},$ $\{b, b1\},$ $\{c, c1\},$ $\{d, d1\}$. Any network with
fewer than four connected components would necessarily place at least
two of $\{a1, b1, c1, d1\}$ in the same component.

This conflict graph also has only one network 
that maximizes its total utility: the $4$-clique plus 4 isolated vertices.  
Both of these networks are stable: 
the $x1$ nodes cannot form any additional edges, and 
no other node would wish to join a pair containing a player she dislikes.
}

Despite these negative results, 
one can still study the quality of $k$-stable networks relative
to these optima as baselines.
We consider the sum of utilities in this section, and the number
of components in the next section.

%
%
%

\subsection{$2$-defections}

For the total utility metric, we can make the
following strong statement: {\em{every}} 
network that maximizes the total utility is 2-stable.
\begin{theorem}\label{thm:2totalwelfareisstable}
In every instance, 
every network that maximizes the total utility is stable.  
Thus, the price of stability for total utility under $2$-deviations is 1.
\end{theorem}

\proof{
Consider a network $G$ that maximizes the total utility.  
We may assume that each component of $G$ is a clique.
Suppose that $G$ is not 2-stable.
Clearly, no player wishes to defect by
simply dropping edges, no two players can form an edge without
dropping any edges (if they could, the network wasn't optimal),
and no two players wish to both drop edges to form a pair.  So we must
consider two players $u$ and $v$ in cliques of size $n_1 \geq n_2$,
respectively, who wish to defect by forming an edge between them while
$v$ drops all of her other edges.  But the resulting total utility
will increase by
\begin{align*}
\frac{1}{2}\left((n_1 + 1 )^2 + (n_2 - 1)^2 - n_1 - n_2 - \left(n_1^2 + n_2^2 - n_1 - n_2\right)\right) \\
= n_1 - n_2 + 1,
\end{align*}
which is strictly greater than 0 for any $n_1 \geq n_2$, so we have arrived at a contradiction, and no player wishes to defect.
}

\begin{figure}
\begin{center}
\includegraphics[height=5cm]{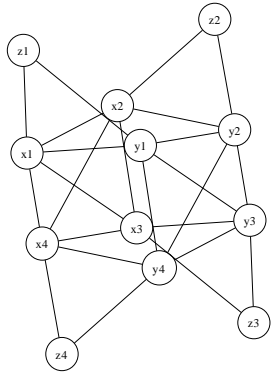}
\end{center}
\caption{
Depicted edges $(i,j)$ represent $u_{ij} = 1$; absent edges have $u_{ij} = -\infty$.}
\label{fig:distinct-stable}
\end{figure}
 
We now show that the price of anarchy is strictly greater than 1
for both the sum of utilities and the number of components.
\begin{theorem}
There exist stable networks that 
neither maximize 
total utility nor minimize the number of connected components.
Hence the price of anarchy for both measures is $>1$.
\end{theorem}
\proof{
  In Figure \ref{fig:distinct-stable}, the minimum number of connected components
  (of four)  is achieved by the partition $\{x1, y1, z1\};$ $\{x2,
  y2, z2\};$ $\{x3, y3, z3\};$ $\{x4, y4,z4\}$.  The maximum total
  utility (value twelve) is achieved by the partition $\{x1, x2,
  x3, x4\};$\\ $\{y1, y2, y3, y4\};$ $\{z1\}; \{z2\}; \{z3\}; \{z4\}$.
  But there is another stable network, with value ten and with five
  connected components: $\{x1, x2, x3, x4\};$ $\{y1, z1\}; \{y2, z2\};
  \{y3, z3\}; \{y4, z4\}$.  This network is stable: no player in a
  pair can entice a player in the group of four to drop all her links,
  none wishes to join the group of four otherwise.  Similarly, no two
  players in different pairs wish to defect.
}

In fact, the price of anarchy for total welfare is much worse.
\begin{theorem}
The price of anarchy for total welfare for $k=2$ is at least $n/2$.
\end{theorem}
\proof{
Consider a conflict graph with two cliques of size $n/2$ with a matching between them except on one pair of unmatched vertices.  The matching is $2$-stable, but the two cliques optimize total welfare.
}

\subsection{$k$-deviations}

We now turn to the problem of characterizing the welfare
properties of stable networks under the more general $k$-defection model.

\paragraph{Price of Stability for total welfare}
Unlike for $k =2$, for larger $k$, not every network maximizing total welfare is stable.

\begin{observation}
For $k=3$, Figure \ref{fig:k3plus3} demonstrates an 
network that maximizes total welfare but is not stable.
\end{observation}

Despite this, for $k=3$, there is 
always a $k$-stable network maximizing total welfare. 
\begin{theorem}
The price of stability for total welfare is $1$ for $k=3$.
\end{theorem}
\proof{
Consider some network maximizing total welfare, and suppose it is not $3$-stable.  We know from the proof of Theorem \ref{thm:2totalwelfareisstable} that there is no defection where fewer than $3$ players participate.  So we will consider every type of $3$-player defection, and perform defections until no more exist.  This will not cycle, because, as we show below, each possible defection strictly increases the potential function that is the sum of the cubes of the connected component sizes.
\begin{packed_itemize}
\item If the three players defect to form a clique and all drop all existing edges, the new total utility is $9 + a^2 + b^2 + c^2$, and the previous utility was $(a + 1)^2 + (b + 1)^2 + (c + 1)^2$, for $a, b, c, \in  \{0, 1\}$.  Thus the new utility is at least the old utility, so this defection will result in a new network maximizing total welfare.  The potential function strictly  increases.
\item Suppose two players drop all existing edges to join a third.  Then if the original total utility was $a^2 + b^2 + c^2$, the new value is $(a - 1)^2 + (b - 1)^2 + (c + 2)^2$.  Since $c + 1 \geq a$ and $c + 1 \geq b$, this defection results in a strict improvement in total utility (of at least two), which is a contradiction.
\item Suppose two players leave the same group to join the third player's group.
If the total utility of the initial network was $a^2 + b^2$, the new total utility is $(a + 2)^2 + (b - 2)^2$, for a strict increase of at least 4, which is a contradiction.
\end{packed_itemize}
Note that we need not consider defections in which more than one player retains links to her current clique, since in that case there would exist a two player defection where these two players form an edge between them.
}

\begin{figure}
\begin{center}
\includegraphics[width=5cm]{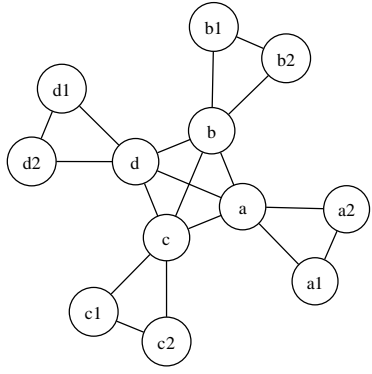}
\end{center}
\caption{Depicted edges $(i,j)$ represent $u_{ij} = 1$; absent edges have $u_{ij} = -\infty$.}\label{fig:k4plustriangles}
\end{figure}

\begin{theorem}
The price of stability for total welfare, when $k \geq 4$, 
is strictly greater than $1$.
\label{thm:k-stab-pos-total-util}
\end{theorem}
\proof{
In Figure \ref{fig:k4plustriangles}, the only
$4$-stable network has components corresponding to 
the $K_4$ plus the four pairs, but this has
lower total utility than the four triangles.  

}

The ratio of $6/5$ implied by the proof of 
Theorem~\ref{thm:k-stab-pos-total-util}
is the strongest lower bound on the price of stability for total welfare that
we know of for any $k$; it is an interesting open question to find the 
correct asymptotic bound for this objective function.

\xhdr{Price of Anarchy for Total Utility}
The worst $k$-stable network can have a 
factor $\frac{n/k - 1}{k -1}$ smaller total utility than is optimal.

\begin{theorem}
The price of anarchy for total welfare 
for $k > 2$ is $\Omega(\frac{n/k-1}{k-1})$.
\end{theorem}
\proof{
Again, consider the graph with $n/k$ rows of elements, with each row forming a clique and each column forming a clique. As before, the columns are stable under $k$-defections; they give total utility $\frac{1}{2}(nk - n)$, whereas the rows give total utility $\frac{1}{2}(n^2/k - n)$.
}

For sufficiently large $k$, however, every stable network gives 
good total utility.
\begin{theorem} \label{thm:welfare-poa-2}
When there exists a socially optimal network where all cliques are of
size $\leq k$, the worst $k$-stable network has at most a factor
$2$ smaller utility than is optimal.

\end{theorem}

\proof{
Consider a clique of size $c \leq k$ that is present in the socially optimal network, and then consider the utility of each of those players in some $k$-stable network.  At least one of those players must have utility at least $c -1$; otherwise the $c$ players would all strongly prefer to defect to their original clique.  Ignoring this first player, there must be some other player from the clique with utility at least $c - 2$; otherwise the $c -1$ remaining players would all prefer to defect and form a clique.  Similarly, there must be players achieving utilities at least $(c - 3), \ldots, 2, 1, 0$.

The utility that this clique contributes to the social optimum is $\frac{1}{2} c (c - 1)$.  The utility those players must contribute to any stable network is at least $\frac{1}{2}\sum_{i = 1}^c (i - 1) = \frac{1}{4} c (c -1)$ (note that our definition of total utility counts each edge once, not twice).  Thus, the total maximum utility is at most twice that of any stable network.
}

In general, the worst $k$-stable network has at most a factor $\frac{n (n -1)}{(k-1)(n - k/2)}$ smaller utility than is optimal.
\begin{theorem} 
The price of anarchy for total utility is $O(\frac{n (n -1)}{(k-1)(n - k/2)}$.
\end{theorem}
\proof{
As in the proof of Theorem \ref{thm:welfare-poa-2}, consider a clique of size $c$ that is present in the socially optimal network.  It contributes $\frac{1}{2}c (c-1)$ to the social optimum, and its constituents must contribute at least $\frac{1}{2} (0 +  1+  \ldots + (k - 2) + (k-1)+ (k-1)+\ldots+ (k -1) = \frac{1}{2}(k -1) (c - k/2)$.  The ratio of the sums of these clique contributions is maximized when we consider a single clique of size $n$.
}

\section{Social Welfare: Number of Components}
\label{sec:components}

We now consider the quality of stable networks with respect to 
the number of components they contain; the ideal outcome is 
to have a number of components close to $\chi(H)$, the chromatic
number of the conflict graph $H$.

\subsection{$2$-stability}

There always exists a 2-stable network 
minimizing the number of connected components.  
\begin{theorem}
In any instance,
there exists a 2-stable network with a number of components
equal to $\chi(H)$.
Thus the price of stability for the number of components
under $2$-deviations is 1.
\end{theorem}

\proof{
Given an instance, and a partition $\Pi$ of the nodes into $\chi(H)$
sets, each of which is independent in $H$, we first build a network
$G$ by placing a clique on each set in $\Pi$, with no other edges between.

Suppose that $G$ is not 2-stable.  It cannot
be the case that there exists a single player who wishes to defect by
dropping edges, since by definition $G$ does not
place any node in a clique with players it dislikes.  It also
cannot be the case that two players wish to defect by forming an edge
between them without dropping any edges, since the resulting
network would have a number of color
classes strictly less than $\chi(H)$.
Nor would both
players wish to drop edges, since their resulting clique would contain
only two players.  So, finally, consider the case where two nodes
wish to defect by forming an edge between them while one drops all its
edges.  Next, form all edges between the defecting node and its new
group members. This results in a new network that also minimizes
the number of components (or reduces, it, if the smaller clique was of
size one), and the defection increases the potential function
corresponding to the sum of the squared group sizes.  Thus we can
allow players to repeatedly defect in this manner until no defections
are available, and the procedure (which started with an arbitrary
network minimizing the number of components) will yield a
{\em{stable}} network minimizing the number of components.
}

\subsection{$k$-stability}

\begin{figure}
\begin{center}
\includegraphics[width=5cm]{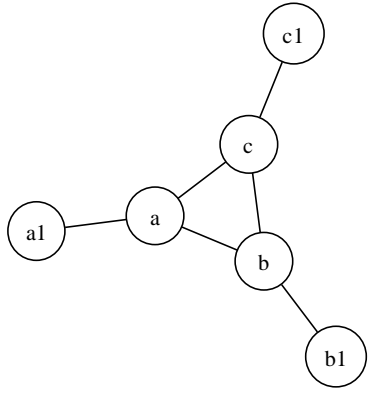}
\end{center}
\caption{Depicted edges $(i,j)$ represent $u_{ij} = 1$; absent edges have $u_{ij} = -\infty$.}\label{fig:k3plus3}
\end{figure}

\xhdr{Price of Stability for Number of Components}
\begin{theorem}
For $k=3$, the price of stability for the number of components is $> 1$.
\end{theorem}
\proof{
Figure \ref{fig:k3plus3} shows a conflict graph
with chromatic number 3, but every 
$k$-stable network for this instance has more than 3 components.
}

\begin{theorem}
The price of stability for the number of components is $O(\log n)$,
for any $k$.
\end{theorem}
\proof{
Algorithm~\ref{alg:undir-def3} is in fact performing the
greedy set-cover algorithm on the set system of independent sets of $H$.
It produces a network $G$ that is $k$-stable for all $k$,
and by the approximation properties of the greedy set-cover algorithm,
it produces a number of components that is at most $\ln s$ times
larger than $\chi(H)$, where $s \leq n$ 
is the maximum independent set in $H$.
}

We now show a matching asymptotic lower bound on the price of stability
when $k = n$.

\begin{theorem}
The price of stability for $n$-defections is  $\Omega(\log n)$.
\end{theorem}
\proof{
Define the graph $B_n$
to have nodes $x_1, \ldots, x_n, y_1, \ldots, y_n$,
with edges $(x_i,y_j)$ and $(x_j,y_i)$ for each pair $(i,j)$ with $j \leq i/3$.
We define an instance with symmetric utilities in $\{-\infty,1\}$
by giving the edges of $B_n$ weight $-\infty$, and all other pairs
of nodes weight $1$.
Since $B_n$ is bipartite, we have $\chi(B_n) = 2$.

We claim that
$B_n$ has a unique maximum independent set $S_1$, equal to
$\{x_i, y_i : i > n/3\}$.
To see why $S_1$ is the unique maximum independent set, consider any
other independent set $R$.
Let $a = \max \{i : x_i \in R\}$ and
$b = \max \{j : y_j \in R\}$.
Now, since $x_a \in R$, we cannot have $y_j \in R$ for any $j \leq a/3$;
and since $y_b \in R$, we cannot have $x_i \in R$ for any $i \leq b/3$.
Thus, $|R| \leq (a - b/3) + (b - a/3) = 2(a + b)/3$.
Now, if $a = b = n$, then $R \subseteq S_1$; and if
$\max(a,b) < n$, then
$|R| < 2(n+n)/3 = 4n/3 = |S_1|$.
Thus, $S_1$ is the unique maximum independent set in $B_n$.

Next, we see that in any stable network $G_n$ 
for the instance defined by $B_n$,
we must have $S_1$ as one of the components;
otherwise, the nodes in $S_1$ could defect and
form a clique on themselves.
But since $B_n - S_1 = B_{n/3}$, we can proceed inductively.
$B_n - S_1$ has a unique maximum independent set $S_2$, equal to
$\{x_i, y_i : n/9 < i \leq n/3\}$.
Since $S_1$ is a component in any stable partition of $B_n$, it follows
that $S_2$ must also be a component in any stable network for 
the instance defined by $B_n$:
otherwise, all nodes in $S_2$ will belong to distinct components
in $B_n - S_1$, with each component an independent set of $B_n - S_1$, and so
they could all improve their utility by defecting to form a clique
on themselves.
Now define $S_{k} = \{x_i, y_i : n/3^k < i \leq n/3^{k-1}\}$.
Continuing by induction, the set $S_k$ is the unique maximum independent set
in the graph $B_n - (\cup_{i=1}^{k-1} S_i) = B_{n/3^{k-1}}$, and
must be a component in any stable network for
the instance defined by $B_n$.
But this implies that any stable network for the instance defined
by $B_n$ must have $\Omega(\log n)$ components.
}

\xhdr{Price of Anarchy for Number of Components}
The worst $k$-stable network 
can have a factor $n/k^2$ more connected components than $\chi(H)$.
\begin{theorem}
The price of anarchy for number of components is  $\Omega(n/k^k)$.
\end{theorem}
\proof{
Consider a graph with $n/k$ rows of elements, with each row forming a
clique and each column forming a clique.  Under size-$k$ defections,
the columns are a stable network, but the chromatic number $k$ is
achieved by the rows.
}

\section{Generalizations of the model}
\label{sec:general}

As noted in the introduction, 
there are several aspects of our model one might consider varying, including
\begin{packed_itemize}
\item Are the values $u_{ij}$ {\em{symmetric}} ($u_{ij} = u_{ji} \, \forall \, i, j$) or {\em{asymmetric}}?  
\item What utility values are allowed?
\end{packed_itemize}
Unfortunately, simple versions of both generalizations result 
in instances for which there is no stable network.

\subsection{Asymmetric Cost Functions}

\begin{theorem}
Under asymmetric preferences, 2-stable networks need not exist
even when all utilities are in $\{-\infty,1\}$.
\end{theorem}
\proof{
Consider a set of four nodes
$x, v_1, v_2, v_3$ and 
$u_{v_1v_2} = u_{v_2v_3} = u_{v_3 v_1} = -\infty$; all other utilities
are $1$.  
In any 2-stable network, none of the nodes $v_i$ can be in a component 
together.
Now, who else can be in a component with $x$?
If $x$ were in a singleton connected component, this would not be stable,
since forming an edge with one of the nodes
$v_i$ node would be an improving defection.
If $x$ were in the same component as (without loss of generality)
$v_1$, then nodes $v_2$ and $x$ would want to gossip, forming a deviation.
Thus, such a network would not be 2-stable either.
}

\subsection{General Symmetric Weights}

As we saw in the introduction, 
stable networks need not exist with symmetric utilities
that can take general values.
In fact, stable networks may not exist 
even in a very mild generalization of our basic model with
symmetric utilities in $\{-\infty,1\}$.
\begin{theorem}
In the model with all
$u_{ij} \in \{-\infty, +1, +c\}$ for $c > n$, stable networks need not exist.
\label{thm:np-hard-matching-structure}
\end{theorem}
\proof{
Suppose we have four nodes $w_1, w_2, m_1, m_2$ with utilities
$u_{w_1m_1} = c; u_{w_2m_2} = c; u_{m_1m_2} = -\infty$ 
and all other utilities equal to $1$.  
Any stable network must put the pairs with utility $c$ in two 
distinct components; but then 
$w_1$ and $w_2$ will wish to defect by forming the edge
between them.  
}

One might ask whether the simple techniques we used in the 
case of symmetric utilities in $\{-\infty, +1\}$ can be generalized to
find stable networks under generalized weights, when such networks
exist.  Unfortunately, this is not the case.  First, repeated
formation of maximum-size cliques does not result in a stable
network in the generalized weight setting setting, since it may place
a player in an earlier clique that is worse for her (but where she
increases its maximum value).  Second, note that sequential improving
moves by single players from one clique to another always increase
the potential function \[\sum_i \sum_{j \in C_G(i)}
u_{ij},\] where $C_G(i)$ denotes the component of $i$ in the current network $G$. The potential function is bounded by the sum of the positive
$u$ values.  However, a chain of sequential improving moves need not
arrive at a stable network, because players might still wish to make
other types of deviations.

Finally, we will show that determining whether an instance 
has a stable network is intractable, 
even under a very mild generalization of the possible utilities.
We represent the utilities using a weighted complete graph $W$, 
in which the weights on the edges of $W$ define the utilities.
The nodes of $W$ are
$x_1, x_2, \ldots, x_n, y_1, y_2, \ldots, y_n$;
there is an edge of weight $c > n$ between each pair $(x_i, y_i)$;
there are edges of weight $-\infty$ between certain pairs $(x_i,x_j)$ and
$(y_k,y_{\ell})$; and there edges of weight $1$ between all other pairs.
We call this a {\em $(-\infty,+1,+c)$-instance with a matching structure}.
\begin{theorem}
For $c > n$, the problem of determining whether a 
{\em $(-\infty,+1,+c)$-instance with a matching structure}
has a stable network is NP-complete.
\end{theorem}
\proof{
We define a set of related problems for the purposes of the reduction.
\begin{packed_itemize}
\item The first is the {\em 3-coloring problem}: 
given a graph $H$, determining whether it is 3-colorable.
\item The second is a problem 
we'll call {\em 3-coloring of a triangle-partitioned graph} (3CTPG).  
In this problem, we are given a graph $H$ together with a
partition of its nodes into triples, each of which induce a triangle in $H$,
and we want to know whether $H$ is 3-colorable.
\item 
The third problem is {\em Stable Coloring of Bichromatic Graphs} (SCBG).
In this problem, we are given a graph $K$ in which each edge
is colored either red or blue.  We allow $K$ to have parallel edges
of different colors.  We want to partition $K$ into 
independent sets $\{S_i\}$ with the property that if $(v,w)$ is an edge of $K$,
with $v \in S_i$ and $w \in S_j$, then 
at least three of the following four kinds of edges are present:
\begin{packed_itemize}
\item[(i)] A red edge from $v$ to a node in $S_j$;
\item[(ii)] A blue edge from $v$ to a node in $S_j$;
\item[(iii)] A red edge from $w$ to a node in $S_i$;
\item[(iv)] A blue edge from $w$ to a node in $S_i$.
\end{packed_itemize}
Such a partition will be called a {\em stable coloring} of $K$.
The problem is to determine whether $K$ has a stable coloring.
\end{packed_itemize}

\xhdr{NP-Completeness of 3CTPG}
We claim that 3CTPG is NP-complete, by a reduction from (standard)
3-coloring.  Given a graph $H$ for which we want to determine
3-colorability, we construct a graph $H'$ as follows:
for each node $v \in V(H)$, we add new nodes $v'$ and $v''$, with
new edges $(v,v'), (v,v'')$, and $(v',v'')$.
We then present $H'$ together with the sets
$\{\{v, v', v''\} : v \in V(H)\}$ as an instance of 3CTPG.
Now, if $H'$ is 3-colorable, then we can use the 3-coloring of
$V(H) \subseteq V(H')$ as a 3-coloring of $H$.
Conversely, if $H$ is 3-colorable, we can extend this to a 
3-coloring of $H'$ by coloring each $v'$ and $v''$ with the two
colors not used for $v$.
Thus $H$ is 3-colorable if and only if $H'$ is, and hence
3CTPG is NP-complete.

\xhdr{NP-Completeness of SCBG}
We next show that SCBG is NP-complete, by a reduction from 3CTPG.
Suppose we are given 
an instance of 3CTPG, consisting of a graph $H = (V,E)$ and a 
partition $\Pi$ of the nodes into sets of size three.
For a node $v \in V$, we let $\pi(v)$ denote the partition $v$ belongs to.
We construct
an equivalent instance of SCBG as follows, consisting of 
a red-blue-colored graph $K$.
For each triangle in $\Pi$, we create a triangle of parallel red
and blue edges on these nodes in $K$.
For all other edges of $H$, we add only a blue edge to $K$.
Note, crucially, that red edges thus only appear in the collection
of disjoint triangles defined by $\Pi$.

Now we claim that $K$ has a stable coloring if and only if 
$H$ has a 3-coloring.
First, suppose that $H$ has a three-coloring, with color classes
$A$, $B$, and $C$.  
Then we use this
same partition of $K$ into three independent sets.
Clearly, for each triangle in $\Pi$, one node goes in each of $A$, $B$, 
and $C$.
As a result, for each pair of nodes $v, w$ in $K$ belonging to different 
color classes, all four types of edges (i)-(iv) are present, since
the other two members of $\pi(v)$ belong to the two color classes $v$
is not in, and $v$ has both red and blue edges to them; and likewise for $w$.
Thus, this is a stable coloring of $K$.

Conversely, suppose that $K$ has a stable coloring.
We first claim that this coloring must 
consist of at most three non-empty independent sets.
Indeed, suppose that this coloring included at least four non-empty independent
sets.
Consider a node $v$ in one of the independent sets $A$, and let 
$\pi(v) = \{v, v', v''\}$ with $v' \in B$ and $v'' \in C$.
Now, since we are assuming there are at least four non-empty independent sets,
let $D$ be another non-empty independent set containing a node 
$w \not\in \pi(v)$.
At least one of $A$, $B$, or $C$ contains no node of $\pi(w)$; suppose
(by symmetry) that it is $C$.
Then $v''$ and $w$ belong to different independent sets in the coloring,
$v''$ has no red edge to any node in $D$, and $w$ has no red edge
to any node in $C$; this contradicts the stability of the coloring.
It follows that the coloring must consist of at most three non-empty
independent sets.
Consequently, the stable coloring of $K$ is also a 3-coloring of $H$;
since this completes the converse direction, we've shown that
$K$ has a stable coloring if and only if $H$ has a 3-coloring.

\xhdr{NP-Completeness of determining whether
a $(-\infty,+1,+c)$-instance with a matching structure has a stable network}
This is the final step, showing that our original problem is NP-complete.
We reduce from SCBG.

Suppose we are given an instance of SCBG, consisting of a 
graph $K = (V,E)$ with each edge colored red or blue.
We construct a $(-\infty,+1,+c)$-instance with a matching structure as
follows, using a weighted complete graph to encode the utilities.
For each $v \in V$, we create two nodes $x_v$ and $y_v$ in $W$, and
we join them by an edge of weight $c$.
Then, 
for each $(v,w) \in E$ colored red, we create an edge $(x_v,x_w)$ 
of weight $-\infty$;
for each $(v,w) \in E$ colored blue, we create an edge $(y_v,y_w)$
of weight $-\infty$.
We include edge of weight $1$ between all other pairs of nodes in $W$.
Now we claim that $K$ has a stable coloring if and only if $W$
has a stable network.

We prove the two directions of this as follows.
First, if there is a stable network $G$ for the instance
$W$, then for each $v$,
the nodes $x_v$ and $y_v$ must be in the same component $S_a$ of $G$, since 
otherwise (by the fact that $c > n$) they would have an incentive
to drop all the edges to their current sets and form an edge
between each other.
We define a subset $S_a' \subseteq V(K)$ consisting of all $v$
for which $\{x_v,y_v\} \subseteq S_a$, and we claim that these sets
$\{S_a'\}$ form a stable coloring of $K$.
Indeed, suppose there existed nodes $v \in S_a'$ and $w \in S_b'$
for which two of the four types of edges (i)-(iv) were not present.
Then this would imply that one of $x_v$ or $y_v$ would be able to 
gossip with one of $x_w$ or $y_w$ without either of them receiving
a negative utility from information spreading into the other's component.
This contradicts the stability of the network $G$ for the instance $W$.

For the converse direction, we must show that if there is a stable
coloring of $K$, into sets $\{S_a'\}$, then there is a stable 
network $G$ for the instance $W$.
For this, we define a set $S_a$ containing both $x_v$ and $y_v$, for
each $v \in S_a$, and include in $G$ a clique of edges on $S_a$.
We put no other edges in $G$.
Since $S_a'$ is an independent set, $S_a$ has no internal $-\infty$ edges.
Also, suppose a node in $z_v \in S_a$ were able to gossip with a node 
in $z_w \in S_b$, without either of these nodes dropping any edges,
where $z_v$ denotes one of the nodes $x_v$ or $y_v$, and
$z_w$ likewise denotes one of the nodes $x_w$ or $y_w$.
Then the corresponding nodes $v$ and $w$ in $K$ would each
lack at least one color of edge into the other's set, 
contradicting the stability of the coloring of $K$.
But neither $z_v$ nor $z_w$ will 
be able to increase utility if they drop edges:
since the sets $S_a$ and $S_b$ are cliques, the only ways that
$v$ and $w$ can break paths to any other nodes in $S_a$ or $S_b$
involve breaking their edges to the nodes to whom they're connected
by edges of weight $c$, which would result in a net loss of utility.
Thus, no nodes have an incentive to change their connections in this partition 
of $G$, and so it is a stable network.

This shows that $K$ has a stable coloring if and only if there
is a stable network for the instance $W$.
and hence establishes the NP-completeness
of our original problem.
}

\bibliographystyle{plain}
\bibliography{gossip,gossip2}

\begin{thebibliography}{10}

\bibitem{adams-friendship}
Rebecca~G. Adams and Rosemary Blieszner.
\newblock An integrative conceptual framework for friendship research.
\newblock {\em Journal of Social and Personal Relationships}, 11(2):163--184,
  May 1994.

\bibitem{aries-friendship}
Elizabeth~J. Aries and Fern~L. Johnson.
\newblock Close friendship in adulthood: Conversational content between
  same-sex friends.
\newblock {\em Sex Roles: A Journal of Research}, 9(12):1183--1195, December
  1983.

\bibitem{BKS01}
S.~Banerjee, H.~Konishi, and T.~S{\"o}nmez.
\newblock {Core in a simple coalition formation game}.
\newblock {\em Social Choice and Welfare}, 18(1):135--153, 2001.

\bibitem{BG07}
S.~Barber{\`a} and A.~Gerber.
\newblock {A note on the impossibility of a satisfactory concept of stability
  for coalition formation games}.
\newblock {\em Economics Letters}, 95(1):85--90, 2007.

\bibitem{bazerman-negotiation}
Max Bazerman, Robert Gibbons, Leigh Thompson, and Kathleen Valley.
\newblock Can negotiators outperform game theory?
\newblock In Jennifer~J. Halpern and Robert~N. Stern, editors, {\em Debating
  Rationality: Nonrational aspects of organizational decision-making}, pages
  78--98. Cornell University Press, 1998.

\bibitem{Becker73}
G.S. Becker.
\newblock {A theory of marriage: Part I}.
\newblock {\em Journal of Political economy}, 81(4):813, 1973.

\bibitem{Becker74}
G.S. Becker.
\newblock {A Theory of Marriage: Part II}.
\newblock {\em Journal of Political Economy}, 82(S2):11, 1974.

\bibitem{BJ02}
A.~Bogomolnaia and M.O. Jackson.
\newblock {The stability of hedonic coalition structures}.
\newblock {\em Games and Economic Behavior}, 38(2):201--230, 2002.

\bibitem{BL09}
S.~Br{\^a}nzei and K.~Larson.
\newblock {Coalitional affinity games and the stability gap}.
\newblock In {\em Proceedings of The 8th International Conference on Autonomous
  Agents and Multiagent Systems-Volume 2}, pages 1319--1320. International
  Foundation for Autonomous Agents and Multiagent Systems, 2009.

\bibitem{DP94}
X.~Deng and C.H. Papadimitriou.
\newblock {On the complexity of cooperative solution concepts}.
\newblock {\em Mathematics of Operations Research}, 19(2):257--266, 1994.

\bibitem{DBHS06}
D.~Dimitrov, P.~Borm, R.~Hendrickx, and S.C. Sung.
\newblock {Simple priorities and core stability in hedonic games}.
\newblock {\em Social Choice and Welfare}, 26(2):421--433, 2006.

\bibitem{EW09}
E.~Elkind and M.~Wooldridge.
\newblock {Hedonic coalition nets}.
\newblock In {\em Proceedings of The 8th International Conference on Autonomous
  Agents and Multiagent Systems-Volume 1}, pages 417--424. International
  Foundation for Autonomous Agents and Multiagent Systems, 2009.

\bibitem{gale-shapley-stable-marriage}
D.~Gale and L.~S. Shapley.
\newblock College admissions and the stability of marriage.
\newblock {\em The American Mathematical Monthly}, 69(1):9--15, 1962.

\bibitem{granovetter-weak-ties}
Mark Granovetter.
\newblock The strength of weak ties.
\newblock {\em American Journal of Sociology}, 78:1360--1380, 1973.

\bibitem{jackson-networks-book}
Matthew~O. Jackson.
\newblock {\em Social and Economic Networks}.
\newblock Princeton University Press, 2008.

\bibitem{MW02}
I.~Milchtaich and E.~Winter.
\newblock {Stability and segregation in group formation}.
\newblock {\em Games and Economic Behavior}, 38(2):318--346, 2002.

\bibitem{Muto90}
S.~Muto.
\newblock {Resale-Proofness and coalition-proof Nash Equilibria}.
\newblock {\em Games and Economic Behavior}, 2:337--361, 1990.

\bibitem{NQ91}
M.~Nakayama and L.~Quintas.
\newblock Stable payoffs in resale-proof trades of information.
\newblock {\em Games and Economic Behavior}, 3:339--349, 1991.

\bibitem{reiman-friendship}
Jeffrey~H. Reiman.
\newblock Privacy, intimacy, and personhood.
\newblock {\em Philosophy and Public Affairs}, 6(1):26--44, Autumn 1976.

\end{thebibliography}




\end{document}